\documentclass[11pt]{article}
\usepackage[utf8]{inputenc}
\usepackage[T1]{fontenc}
\usepackage{lmodern}
\usepackage[margin=1in]{geometry}
\usepackage{graphicx}
\usepackage{booktabs}
\usepackage{amsmath,amssymb}
\usepackage{caption}
\usepackage[hidelinks]{hyperref}
\usepackage{microtype}
\usepackage[round]{natbib}
\title{Bitcoin Runs on a Clock: Why Every Price Indicator Dies and the Halving Clock Doesn't}
\author{Josh Molnar\\ \textit{Independent Researcher}}
\date{June 2026}

\begin{document}
\maketitle

\begin{abstract}
Every widely followed Bitcoin cycle indicator (the Pi Cycle Top, MVRV, the Mayer Multiple, the Puell Multiple, the 200-week moving average) called the market's tops and bottoms with striking precision for a decade, then degraded in a common sequence: precise, then early, then silent. We show that this is not a series of independent failures but a single structural phenomenon. Using daily data spanning the four halving epochs (2011--2026; the fourth is still in progress), we document three things. First, the per-cycle maxima of the five major top-calling oscillators decline monotonically across all four epochs while their per-cycle minima end the era higher than they began, so any threshold calibrated on past cycles must eventually stop firing. Second, the predictive power of short-horizon price indicators decays toward zero across epochs and in several cases inverts sign. Third, Bitcoin's \emph{time} structure remains fixed: the three mature-cycle tops occurred 525, 546, and 534 days after their respective halvings, and the three completed bear-market bottoms 406, 364, and 366 days after their tops. These turns are identified retrospectively by a fixed mechanical rule (an all-time high followed by a $\geq$45\% drawdown), not from a real-time forecasting record. Under uniform timing-free nulls, the joint probability of both clusters ranges from $5\times 10^{-6}$ to $1\times 10^{-3}$ across every window choice and inclusion variant we examine, including the variant least favorable to the claim. A harder empirical null, in which block-bootstrapped Bitcoin return paths are processed by the identical turn rule, never reproduces the top-timing regularity under its deterministic construction (zero of 10,000 paths across five block lengths; a conservative variant that grants the null free selection among candidate tops yields 0.10--0.25\%). The same null reveals that the \emph{bottom}-timing regularity is largely intrinsic to the drawdown process itself, reproduced by 31--43\% of null paths: the clock evidence is concentrated in the tops' halving-phase alignment, and we weight our predictions accordingly. A causally fit power law in time-since-genesis (exponent converging to 5.6) is the only signal in the formal inference battery whose direction is stable across every mature epoch, whose long-horizon relationship replicates on an independent price source and on Ethereum with a different exponent, and whose risk-adjusted timing edge over buy-and-hold improves monotonically across all four cycles, turning positive in the current one (a single-window holdout, suggestive rather than decisive). We deliberately rest none of this on per-epoch significance testing: a dependence-preserving rotation null shows that autocorrelation-robust (HAC) inference over-rejects severely for persistent signals at these sample sizes (empirical size up to 0.33 at nominal 0.05), rendering the apparent two-epoch significance of the power-law cell statistically indistinguishable from chance (empirical p = 0.21). We report this diagnostic in full, because it disqualifies not only our own per-epoch p-values but the entire genre of such claims for persistent indicators in this literature. Candidate macro drivers (M2 growth, the yield curve) exhibit the same era-dependent sign instability as the price indicators and are eliminated in a joint horse race; in the 2024--26 episode, liquidity kept expanding while Bitcoin topped on schedule and entered a bear market. The halving clock further outperforms every look-alike four-year cycle we test (the US election cycle, a fixed calendar, and 2,000 random four-year clocks), and the whole boom-bust-recovery path, not just the turns, repeats when cycles are aligned by days since halving. We summarize the asset's state in two time-anchored coordinates, days since halving and power-law deviation, under which Bitcoin's history traces a damped spiral: the amplitude is dying, the clock is not. Re-expressing the clock in its native protocol unit, block height rather than calendar days, holds the tops' null at zero of 10,000 and widens its separation from look-alike clocks, because the halving is an exact 210,000-block period rather than a drifting four-year one; it also recovers partial bottom structure the calendar null misses, though the cycle-shape and volatility overlays do not improve. The framework yields pre-registered, falsifiable predictions: a cycle bottom window of October 5 to November 16, 2026, and a next cycle top window 525 to 546 days after the following halving (provisionally late September to mid-October 2029, pending the halving's exact date).
\end{abstract}


\section{Introduction}

On January 8, 2009, five days after mining Bitcoin's genesis block, Satoshi Nakamoto announced the system's monetary schedule to the Cryptography Mailing List: total circulation would be 21 million coins, distributed to network nodes when they make blocks, "with the amount cut in half every 4 years." The schedule is not a policy that an institution maintains; it is a constant in the protocol, \texttt{SubsidyHalvingInterval = 210000} blocks, fixed at block zero and executable by no one's discretion. It is, to date, the only macroeconomic commitment in financial history that is known with certainty decades in advance.

This paper asks a simple question: \emph{what is the stable coordinate system of Bitcoin's price dynamics, price or time?}

The question is motivated by an empirical puzzle. Over Bitcoin's first decade, a family of price-anchored and on-chain indicators earned reputations as infallible cycle-callers. The Pi Cycle Top indicator marked the December 2013 top to within one day, the December 2017 top to the day, and the April 2021 peak to within one day. MVRV above 3.7 and the Mayer Multiple above 2.4 flagged every major top from 2013 through early 2021 within a month; the Puell Multiple above 4.0 flagged every top through 2017. On the other side, MVRV below 1.0 and the Puell Multiple below 0.5 marked the 2015, 2018, and 2022 bottoms within four weeks. These records were earned out of sample, in real time, across multiple cycles. They were not curve-fit artifacts.

\begin{figure}[htbp]
\centering
\includegraphics[width=\linewidth]{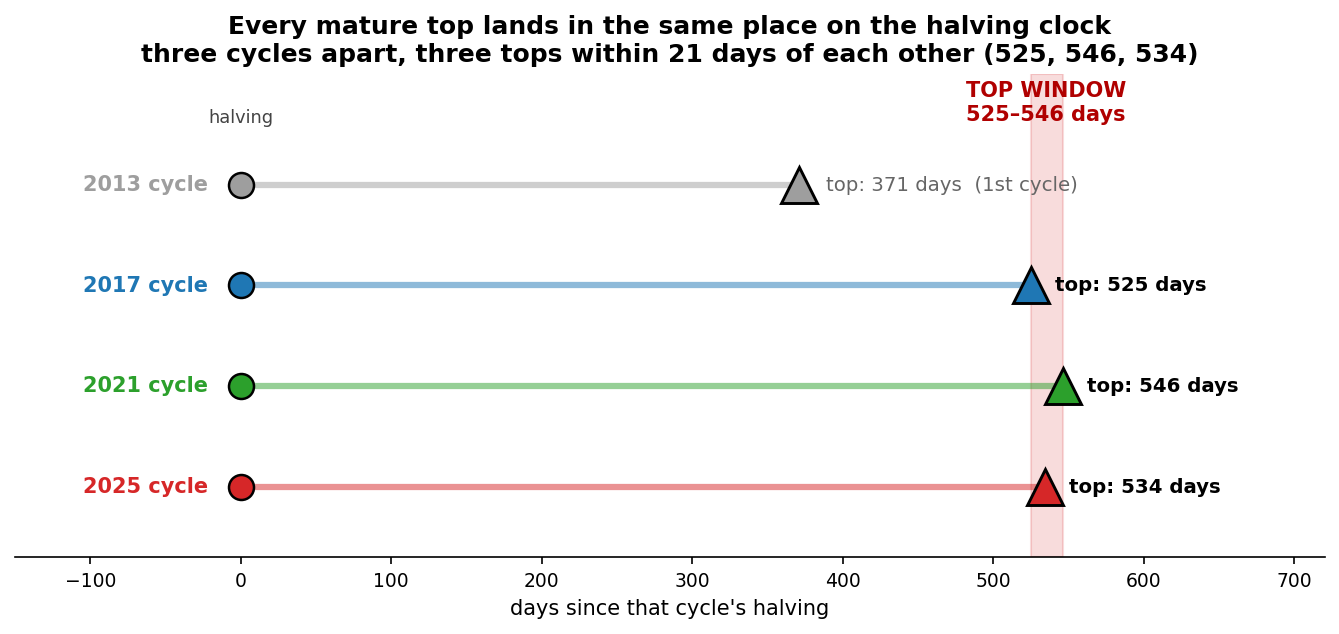}
\caption{Every mature cycle top lands in the same narrow band on the halving clock: 525, 546, and 534 days after its halving, three cycles apart. The 2013 first cycle, still immature, peaked earlier at 371 days. This is the paper's central claim in one picture.}
\label{fig:figC}
\end{figure}

Then, at the October 2025 cycle top, the largest in Bitcoin's history by price, not a single one of them fired. An investor relying on the indicators that had "never missed" received no warning at all. Yet the top itself arrived almost exactly where Bitcoin's previous two mature cycles would have put it: 534 days after the April 2024 halving, against 525 days in 2017 and 546 days in 2021. Three cycles apart, the mature tops land within 21 days of one another when measured on the halving clock (Figure~\ref{fig:figC}).

We argue that these two facts, the simultaneous silence of every price-anchored indicator and the punctuality of the turn in halving-time, are the same fact viewed from two sides. Together they identify Bitcoin's market cycle as a \emph{damped oscillation on a fixed clock}. Each cycle, the oscillation's amplitude shrinks: we show that the per-cycle maximum of every major oscillator declines monotonically across all four epochs (MVRV cycle peaks of 5.88 $\rightarrow$ 4.72 $\rightarrow$ 3.96 $\rightarrow$ 2.74, for example), while per-cycle minima end the era higher than they began. Range compression mechanically silences any fixed threshold calibrated on earlier cycles. "It never missed before" described a shrinking range, not a reliable indicator. The period, by contrast, does not move: in time-since-halving coordinates, the turns cluster within $\pm$2\%.

Our contributions are fivefold:

\begin{enumerate}
\item \textbf{A causal power-law methodology.} We estimate Bitcoin's time trend (price $\approx$ A$\cdot$t$^{n}$, t = days since genesis) by \emph{expanding-window} regression, so that the trend at every date uses only information available on that date. This distinguishes our approach from the discredited Stock-to-Flow tradition, in which the trend is fit through the very future it is then claimed to predict. The causal exponent converges to n $\approx$ 5.6.
\item \textbf{A measurement of cross-cycle indicator decay.} Using rank information coefficients with autocorrelation-robust inference (Newey--West and circular block bootstrap), we show that short-horizon price indicators decay to statistical noise across epochs (RSI-14's 30-day IC falls from +0.28 to $-$0.02), that several invert sign, and that the long-horizon power-law deviation is the only signal in the formal inference battery whose relationship to forward returns holds direction in the two most recent cycles.
\item \textbf{The mechanism.} The compression of indicator extremes (top-side maxima declining monotonically across five oscillators in all four epochs; bottom-side minima ending the era higher than they began) explains \emph{why} threshold indicators fail in sequence ordered by threshold aggressiveness, and predicts that recalibration cannot help: a threshold fit to past amplitude is a bet that damping has stopped.
\item \textbf{Elimination of macro confounders.} Money-supply growth and the yield curve exhibit the same era-dependent sign instability as the price indicators; in a joint HAC regression only cycle phase survives; turn-timing in liquidity coordinates is an order of magnitude looser than in halving coordinates; and the 2024--26 episode provides a natural experiment in which the liquidity thesis and the clock disagreed: liquidity expanded while Bitcoin topped on schedule.
\item \textbf{The Satoshi Clock.} We condense the framework into a two-coordinate state description, CLOCK (days since halving) and SPRING (power-law deviation), under which Bitcoin's full history traces a damped spiral, and we pre-register two falsifiable predictions (Section 8).
\end{enumerate}

We are explicit about what we do not claim. With three mature cycles, we cannot establish that the halving \emph{causes} the cycle, and we do not attempt to. Our claim is positional: among all candidate coordinate systems examined (price levels, price-derived oscillators, on-chain ratios, monetary aggregates, the yield curve), the protocol's time structure is the only one whose relationship to Bitcoin's dynamics is stable across eras, and the only one knowable in advance. Whether the clock works through real supply economics, through self-fulfilling coordination on a schedule that every participant can read (a Schelling point in code), or both, is a question we leave open. Under either mechanism, the clock is the load-bearing element.

The remainder of the paper proceeds as follows. Section 2 situates the work. Section 3 describes data. Section 4 develops the methodology, with emphasis on causality discipline. Section 5 presents results in nine parts. Section 6 interprets. Section 7 states limitations. Section 8 pre-registers predictions. Section 9 concludes.

\section{Related Work}

\textbf{Bitcoin price models in time.} Santostasi and Perrenod (2026), in the first peer-reviewed treatment of the model, derive Bitcoin's price as a power law in time since genesis, price $\propto$ $t^{5.6}$9, from network-adoption dynamics and generalized-Metcalfe scaling; Burger (2019) \cite{burger2019} independently described a closely related long-term power-law corridor of growth. Both are descriptive and fit on the full sample, yet our causal estimate (Section 4.1) recovers essentially the same exponent, 5.61, with no look-ahead. PlanB's (2019) \cite{planb2019} Stock-to-Flow model related price to programmed scarcity and gained wide attention before its post-2021 predictions failed by an order of magnitude; Cordeiro (2020) \cite{cordeiro2020} and others criticized its econometrics, in particular the in-sample fit and non-stationarity. Our methodological departure from both is the \emph{causal} (expanding-window) estimation of the time trend: every parameter at date t is computed from data through date t only, so the trend cannot borrow information from the future it is evaluated against. The same discipline is applied to every signal in the study.

\textbf{Market efficiency and adaptation.} Urquhart (2016) \cite{urquhart2016} found Bitcoin returns inefficient but trending toward efficiency; Nadarajah and Chu (2017) \cite{nadarajah2017} and Bariviera (2017) \cite{bariviera2017} refined and extended the finding. Lo's (2004) \cite{lo2004} Adaptive Markets Hypothesis predicts exactly the pattern we measure: profitable regularities decay as capital learns them, at a rate proportional to how easily they are arbitraged. Our contribution to this literature is a \emph{differential} decay measurement. Price-anchored signals decay to noise while a time-anchored structure does not, which suggests that what is being arbitraged away is amplitude, not timing.

\textbf{Halving studies.} Event studies of the halvings (e.g., Meynkhard, 2019 \cite{meynkhard2019}; subsequent work reviewed in MDPI's \emph{JRFM}) find positive but delayed price responses, with predictive power that several authors describe as weakening. Practitioner literature, much of it built on the on-chain valuation tradition popularized by Woo (2017) \cite{woo2017}, widely asserts that the four-year cycle is dead or macro-driven. Our results cut against both views in a specific way: the cycle's \emph{amplitude} response is indeed weakening, consistent with the maturation literature, but its \emph{timing} regularity strengthened. The 2025 top was the most punctual yet.

\textbf{Liquidity narratives.} A large practitioner literature attributes Bitcoin's cycle to global money-supply growth, often citing a roughly 10-week lead of M2 over price. We provide, to our knowledge, the first per-epoch stability test of this relationship, and find it to be a one-era artifact (Section 5.6).

\section{Data}

\textbf{Bitcoin price.} Primary series: Bitstamp BTC/USD daily OHLC, August 18, 2011 to June 10, 2026 (5,411 days), spanning all four halving epochs and including intraday ranges for honest stop/target evaluation. Independent cross-check and long-history close: Coin Metrics community PriceUSD, July 18, 2010 to May 23, 2026 (5,789 days). The two sources agree to a median absolute difference of 0.15\% over 5,393 overlapping days (95th percentile 1.82\%; early-2010s single-exchange prints account for the tail).

\textbf{On-chain.} Coin Metrics community series: CapMVRVCur (MVRV ratio) and IssTotUSD (daily issuance value in USD, the Puell Multiple numerator), full span.

\textbf{Block height.} For the block-coordinate analysis of Section 5.9 we construct a daily date-to-height series from the Coin Metrics community BlkCnt series (blocks mined per UTC day), contiguous from the January 2009 genesis with no calendar gaps, cumulatively summed to an end-of-day height. Two independent ground-truth checks validate it: the four protocol halving heights (210,000 / 420,000 / 630,000 / 840,000) are reached within one day of their known dates, and eleven mempool.space point-lookups at the dated turns agree to a constant sampling offset (median 75 blocks above the noon-sampled lookups, which cancels in every range statistic). The series lags the price data by about eight weeks, ending May 23, 2026; the block-phase battery needs only history through the October 2025 top, and we do not extrapolate the tail.

\textbf{Ethereum.} Coin Metrics PriceUSD, August 8, 2015 to May 23, 2026 (3,942 days), for the cross-asset replication; genesis date 2015-07-30.

\textbf{Macro.} FRED M2SL (US M2, monthly from 2018, quarterly from 2001) and T10Y2Y (10-year minus 2-year Treasury spread, quarterly from 2001), obtained via the TradingView FRED mirror and verified internally: 34 overlapping quarter-ends between the monthly and quarterly M2 series agree exactly. Publication lags are applied before any predictive use, +28 days for M2 and +7 days for the spread. One limitation: the series is US M2 rather than a global aggregate.

\textbf{Epochs and turns.} Epochs are delimited by the halvings: 2012-11-28, 2016-07-09, 2020-05-11, 2024-04-20 (E1--E4; inter-halving gaps 1,319 / 1,402 / 1,440 days). Cycle turns are defined mechanically, with no hand-picking. A \textbf{cycle top} is an all-time high followed by a drawdown of at least 45\% before any new high (the final running peak qualifies "in progress" once the drawdown exceeds 35\%); a \textbf{cycle bottom} is the minimum close between consecutive tops. Applied to the data from 2012 onward, this rule recovers exactly the turns of market folklore: tops April 2013, December 2013, December 2017, April 2021, November 2021, October 2025; bottoms July 2013, January 2015, December 2018, July 2021, November 2022 (on the full span the rule additionally flags four 2010--2011 microcap-era turns, all before the first halving, which enter no analysis). No discretionary dating is involved at any point.

\section{Methodology}

\subsection{The causal power law}

\begin{figure}[htbp]
\centering
\includegraphics[width=\linewidth]{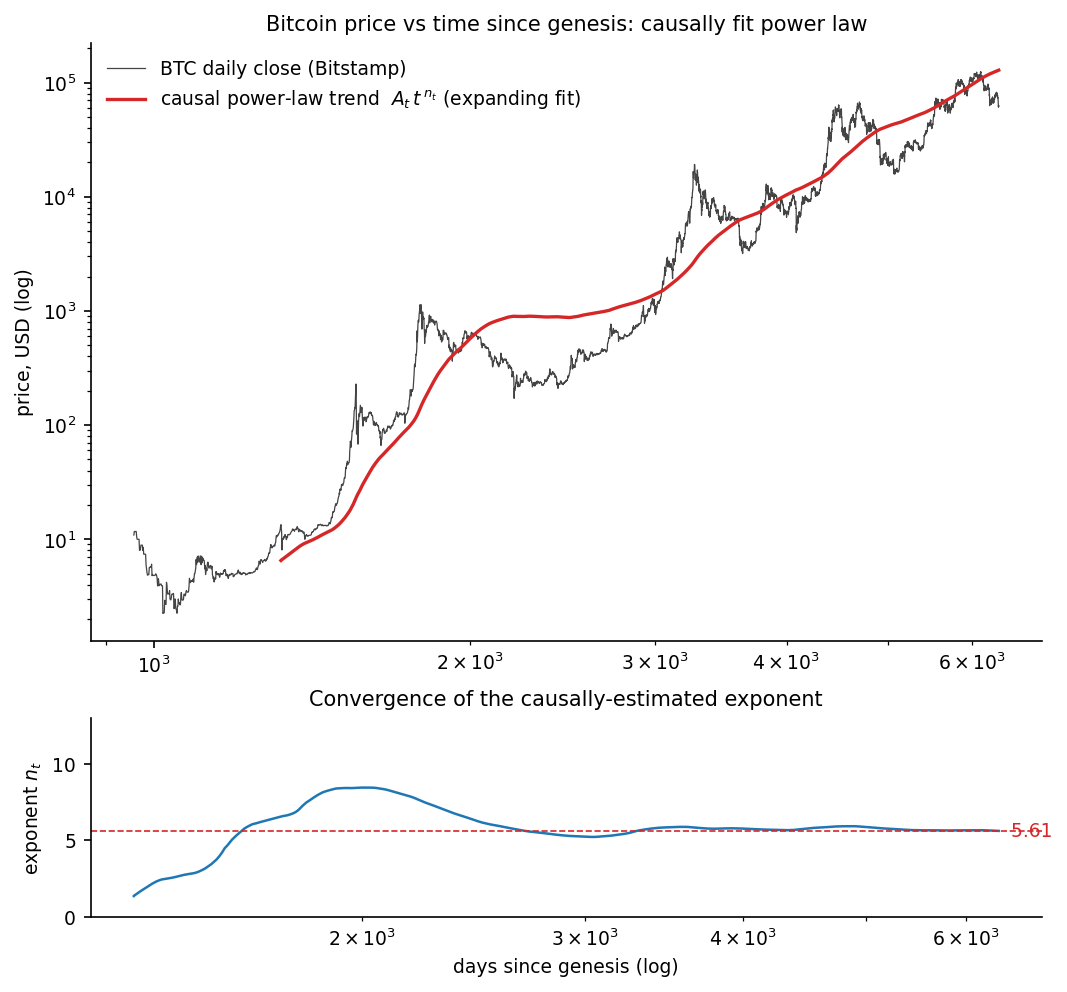}
\caption{Bitcoin price vs time since genesis with the causally fit power-law trend (top) and the convergence of the expanding-window exponent to 5.61 (bottom).}
\label{fig:fig1}
\end{figure}

Let P\_t be the daily close and $\tau$\_t the number of days since the genesis block (January 3, 2009). At each date t we estimate by ordinary least squares, using only observations up to t,

\[ \ln P_{s} = \ln A_{t} + n_{t} \cdot \ln \tau_{s} + \varepsilon_{s},  s \leq t \]

via running sums (exact, O(N)), requiring a minimum of 365 observations. The residual r\_t = ln P\_t $-$ (ln A\_t + n\_t ln $\tau$\_t) measures how far price stands from \emph{its own trend as estimable on that day}. We standardize r\_t by its expanding mean and standard deviation to obtain the deviation score z\_t ("SPRING" below). The fitted exponent stabilizes early and converges to n = 5.61 on Bitstamp data (5.67 on Coin Metrics), consistent with the value near 5.8 reported from full-sample fits (Figure~\ref{fig:fig1}). No parameter anywhere in this paper is estimated with future data. Each signal decides at a bar's close; any trade-structure evaluation enters at the next bar's open.

\subsection{Signal battery}

All signals are causal by construction: RSI-14 (Wilder); the Mayer Multiple (price / 200-day SMA) and its expanding z-score; the Pi Cycle ratio (111-day SMA / 2$\times$350-day SMA); price relative to the 200-week and 2-year moving averages; MVRV; the Puell Multiple (issuance USD / its 365-day mean); the power-law deviation z; and \emph{phase}, defined as days since the most recent halving: a pure time signal containing no price information whatsoever.

\subsection{Predictive evaluation}

For horizon h $\in$ \{30, 90, 180\} days we compute the Spearman information coefficient (IC) between each signal and the forward h-day return, within each epoch. Because overlapping h-day returns are MA(h$-$1) autocorrelated, naive significance overstates by roughly $\sqrt{\,}$h. We therefore report (i) Newey--West (NW) t-statistics from rank-on-rank regression with Bartlett lag h, and (ii) circular moving-block bootstrap confidence intervals (block length min(h, n/5), B = 2,000), and accept a result only when both agree. For the head-to-head claim that one signal beats another, we additionally bootstrap the \emph{paired} difference |IC\_A| $-$ |IC\_B| within identical resamples.

\textbf{Multiplicity and calibration: what the grid cannot support.} HAC corrects autocorrelation, not the garden of forking paths, and at these sample sizes it does not even fully correct autocorrelation. Two diagnostics govern what the 3 signals $\times$ 3 horizons $\times$ 4 epochs = 36-cell grid can carry (the wider nine-indicator battery of Section 5.4 is descriptive only). \emph{(i) False-discovery control:} under Benjamini--Hochberg across all 36 cells, no single cell survives at q = 0.05 or q = 0.10. The smallest individual p in the grid belongs to RSI in its 2016--2020 golden era (p = 0.015), a cell whose signal subsequently died. \emph{(ii) An empirical replication null:} the apparent strength of the power-law cell is that it is the only one "significant" with consistent sign in both of the two most recent epochs. We test how often that configuration arises by chance with a dependence-preserving rotation null: within each epoch, the forward-return series is circularly rotated against the signal block by a random offset (one shared offset per epoch per replicate), preserving every series' autocorrelation, the cross-signal correlations, and the overlap structure, while destroying signal-to-future alignment; 2,000 replicates. The result is disqualifying for per-epoch inference. The NW statistic rejects at empirical rates of 0.33 (E3) and 0.17 (E4) per cell against a nominal 0.05, because persistent signals run against overlapping returns generate spurious "significance" constantly, and the probability that some cell double-hits with consistent sign is 0.21. The observed double-hit of the power-law cell is therefore statistically indistinguishable from persistence artifact. We accordingly use the IC tables in this paper \emph{descriptively}, for sign stability and monotone cross-epoch trends, which are model-free statements about the realized sample, and rest economic claims on out-of-sample structure (Sections 5.4--5.5) and on the turn-timing test of Section 5.1, whose discrete-event null does not involve persistent-regressor correlation. We flag the calibration failure prominently because it applies with equal force to any published claim of per-epoch indicator significance on persistent crypto signals.

\subsection{Economic evaluation}

Two designs. (a) A discrete trade structure, identical across signals, that buys when a signal marks price "cheap" (z \textless{} $-$1, or RSI \textless{} 30), exits on reversion to trend (z $\geq$ 0 / RSI $\geq$ 55), with a 35\% catastrophe stop evaluated intrabar against daily lows and fees included. (b) An all-days long/flat exposure (long when below trend or oversold, causal one-day lag), whose annualized Sharpe ratio is compared per epoch against always-long buy-and-hold. Buy-and-hold is the appropriate null for a 15-year secular bull.

\subsection{The clock's null test, and what it can and cannot mean}

The clock's regularity consists of six retrospectively identified turns: three mature-cycle tops at 525/546/534 days after their halvings (range 21 days) and three completed bottoms at 406/364/366 days after their tops (range 42 days). Three disclosures govern the interpretation of everything in this subsection.

\emph{First, the turns are ex post.} They are dated by the mechanical rule of Section 3, applied to history; no one called them publicly in advance. The resulting probabilities therefore quantify how surprising the retrospective regularity is under a timing-free null. They are descriptive, and cannot substitute for a forecasting record. The pre-registered windows of Section 8 constitute the first real-time test of the clock.

\emph{Second, the test postdates the observation.} Like any analysis of a noticed pattern, the null test was designed after the clustering was seen. We mitigate the resulting window-choice degrees of freedom by fixing the statistic (the observed range) and reporting a full sensitivity grid rather than a chosen window; we cannot eliminate post-hoc-ness itself.

\emph{Third, the inclusion rule is asymmetric and we report the unfavorable variant.} Our headline excludes the first cycle's top (371 days; the documented warmup epoch) while including its bottom (406 days after the top). Because the warmup rationale concerns the power-law fit rather than top timing per se, we also report the variant with the E1 top included (four tops, range 175 days), the least favorable defensible specification.

The null: each top uniform on a window of width W\_top days after its halving; each bottom uniform on W\_bot days after its top. The two gap families are disjoint random variables under this null, so the joint probability is their product by construction. P(range $\leq$ r) for n uniform points is n(r/W)\textasciicircum{}(n$-$1) $-$ (n$-$1)(r/W)\textasciicircum{}n, exact; Monte Carlo ($2\times 10^{6}$ draws) agrees within sampling error.

\textbf{Table 0. The full variant grid (no cell hidden).}

\begin{table}[htbp]
\centering
\small
\begin{tabular}{lcc}
\hline
Variant & W generous & W conservative \\
\hline
Tops, mature only (n=3, r=21d) & 0.0006 (W=1458) & 0.0026 (W=700) \\
Tops, E1 included (n=4, r=175d) & 0.0063 (W=1458) & 0.0508 (W=700) \\
Bottoms (n=3, r=42d) & 0.0080 (W=800) & 0.0200 (W=500) \\
Joint, headline & $4.9\times 10^{-6}$ & $5.3\times 10^{-5}$ \\
Joint, hostile (E1 top incl.) & $5.0\times 10^{-5}$ & $1.0\times 10^{-3}$ \\
\hline
\end{tabular}
\end{table}

The regularity survives the most hostile defensible specification at p $\approx$ $10^{-3}$ and the headline specification at $10^{-5}$ to $10^{-6}$. We treat these as motivation for the forward test, not as confirmation.

\textbf{The empirical drawdown-process null (the referee's null).} The uniform null is vulnerable to a serious objection: perhaps peaks followed by 45\% drawdowns occur at a roughly four-year cadence naturally in an asset with Bitcoin's volatility structure, halving or no halving. We therefore construct a null that concedes everything except the calendar: circular block-bootstrap of Bitcoin's own daily log returns (block lengths 30/60/90/120/250 days, preserving volatility clustering and drawdown geometry), generating 2,000 synthetic price histories per block length over the same calendar, each processed by the \emph{identical} mechanical turn rule.

The top statistic comes in two constructions. The primary construction is deterministic and contains zero analyst freedom: the cycle top of an epoch is its highest-priced qualifying top. Since every qualifying top is by definition an all-time high, this is mechanically the \emph{last} qualifying top of the epoch. Under this rule November 2021, not April 2021, simply is the 2021 cycle top; no selection exists for anyone to make. The observed range is 21 days either way, because November 2021 is simultaneously the higher-priced and the range-minimizing top. A conservative secondary construction grants the null every selection freedom instead (the range-minimizing top per epoch; the best 3-subset of completed top-to-bottom gaps), bounding the result from above against any accusation that a selection favored us.

Results: synthetic paths produce on average about four qualifying mature-epoch tops, so large drawdowns at roughly this cadence are indeed natural. But under the deterministic construction, \emph{no synthetic path in 10,000} (2,000 per block length, five block lengths) reproduces the observed $\leq$21-day halving-phase cluster; under the conservative construction, 0.10--0.25\% do. The deterministic version is tighter for a clean reason: locked to each epoch's actual peak at a random phase, the null can no longer assemble a tight combination from multiple candidate tops. The same null exposes an honest casualty, but a casualty specific to the anchor. Measured forward from the top, the bottom-gap cluster ($\leq$42 days) is reproduced by 31--43\% of null paths: bear durations near twelve months are largely intrinsic to the drawdown process and carry little independent evidence about the clock. Measured backward from the \emph{next} halving, however, the same three bottoms cluster in a 29-day window that this null reproduces in only 1.5\% of paths, against 39\% for the after-top anchor under identical machinery; the bottoms appear pinned to the upcoming halving more tightly than to anything behind them (as the halving gaps grew across cycles, the bottoms drifted later from the previous halving but held constant relative to the next). We do not promote this to a result. The next-halving anchor was one of several we examined after seeing the data, it rests on three bottoms, and a forward halving is itself a drifting target; we instead pre-register it as a falsifiable expectation for the 2026 bottom (Section 8). The case still rests on the tops, and the cadence is not the surprise; the phase-lock is. This construction was independently reconstructed from scratch by a second analyst on independently pulled data, with corroborating results under both constructions.

\section{Results}

\subsection{The clock's regularity and its null test}

\textbf{Table 1. Every completed cycle turn at mature scale (mechanical definition, ex post).}

\begin{table}[htbp]
\centering
\small
\begin{tabular}{lcccc}
\hline
Turn & Date & Price & Days after halving & Days after its top \\
\hline
Top (E2) & 2017-12-16 & \$19,641 & 525 &  \\
Top (E3) & 2021-11-08 & \$67,542 & 546 &  \\
Top (E4) & 2025-10-06 & \$124,824 & 534 &  \\
Bottom (E1) & 2015-01-14 & \$176 & 777 & 406 \\
Bottom (E2) & 2018-12-15 & \$3,185 & 889 & 364 \\
Bottom (E3) & 2022-11-09 & \$15,758 & 912 & 366 \\
\hline
\end{tabular}
\end{table}

\emph{Prices are the Coin Metrics reference daily close at each mechanically dated turn; it is the dating, not the price level, that supports the day-count claims. For reference, the October 2025 cycle's intraday all-time high was \$126,296 on Coinbase and \$126,272 on Bitstamp.}

Table 1 lists every completed turn at mature scale. The mature tops cluster at 535 $\pm$ 10 days ($\pm$2\%); as a fraction of their varying halving gaps they sit at 0.374 / 0.379 / 0.366 (the last computed against the protocol-expected 1,458-day gap to the next halving). The bottoms cluster at 364--406 days after their tops, in three bear markets of very different character (a chaotic post-mania collapse, an orderly grind, and a leverage cascade). The first cycle's top (371 days) is the documented exception, consistent with every other warmup effect in the study (Section 5.4).

Under the null of Section 4.5, with the three disclosures stated there governing interpretation, the joint probability of both clusters ranges from $4.9\times 10^{-6}$ (headline specification, generous windows) to $1.0\times 10^{-3}$ (the hostile variant with the E1 top included, most conservative windows). Monte Carlo agrees with the exact formula within sampling error, and Table 0 reports every cell of the variant grid. The retrospective regularity is very unlikely under a timing-free null in every defensible specification; whether the clock holds \emph{prospectively} is the subject of Section 8.

\begin{figure}[htbp]
\centering
\includegraphics[width=\linewidth]{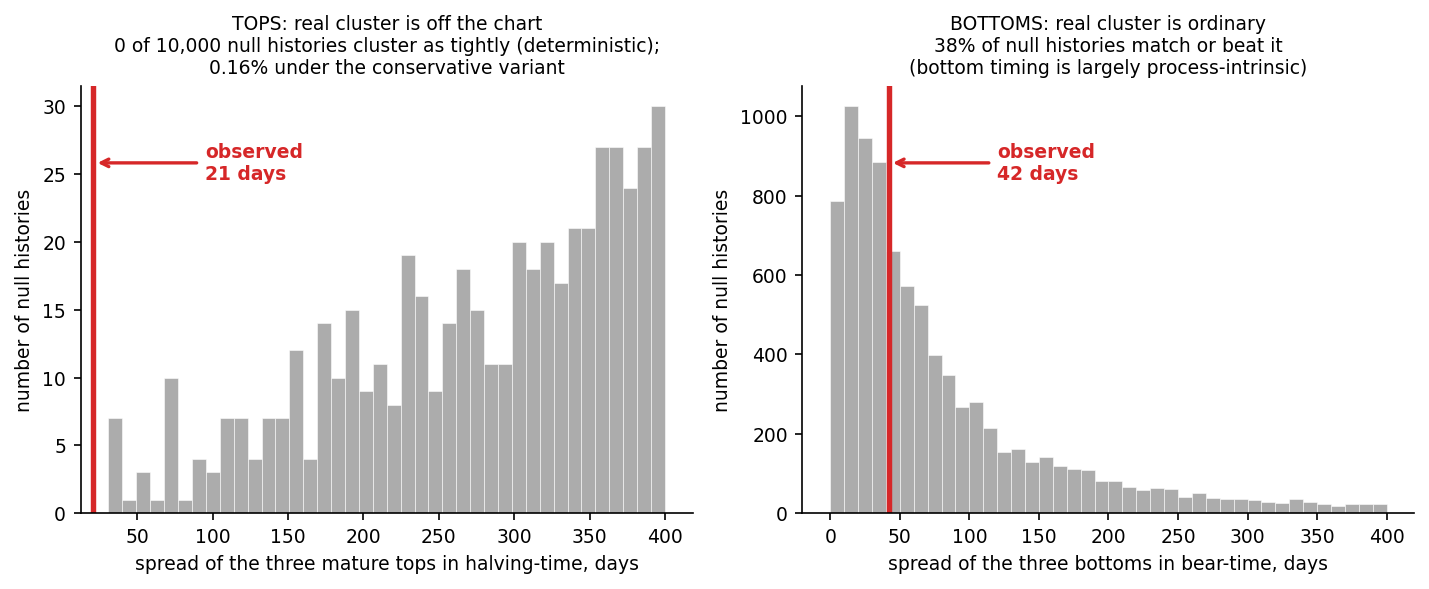}
\caption{The turn-timing null. Left: across 10,000 block-bootstrapped Bitcoin histories run through the identical mechanical turn rule, none clusters its three mature tops as tightly as the observed 21 days (zero under the deterministic construction; 0.16\% under the conservative selection-symmetric variant). Right: the same null reproduces the bottom cluster in 38\% of histories, so bottom timing is largely intrinsic to the drawdown process, not the clock.}
\label{fig:figA}
\end{figure}

The harder empirical drawdown-process null of Section 4.5 makes the asymmetry between the two clusters visible directly (Figure~\ref{fig:figA}): not one of 10,000 block-bootstrapped Bitcoin histories reproduces the tops' 21-day cluster under the deterministic construction, while 38\% reproduce the bottoms' 42-day cluster. This is the single clearest statement of where the clock evidence lives, and it is why we weight the forward predictions toward the tops.

\subsection{The cycle-callers: precise, then early, then silent}

\textbf{Table 2. Canonical published thresholds versus the mechanically dated tops (offset of nearest firing; HIT if within 60 days).}

\begin{table}[htbp]
\centering
\small
\begin{tabular}{lcccccc}
\hline
Trigger & Apr 2013 & Dec 2013 & Dec 2017 & Apr 2021 & Nov 2021 & Oct 2025 \\
\hline
Pi Cycle cross & HIT $-$3d & HIT +1d & HIT +0d & HIT $-$1d & $-$210d & no fire \\
MVRV \textgreater{} 3.7 & HIT $-$21d & HIT $-$26d & HIT $-$10d & HIT $-$31d & $-$240d & no fire \\
Mayer \textgreater{} 2.4 & HIT +25d & HIT +19d & HIT $-$19d & HIT $-$31d & $-$240d & no fire \\
Puell \textgreater{} 4.0 & HIT $-$20d & HIT +22d & HIT $-$19d & no fire & no fire & no fire \\
Price \textgreater{} 5$\times$2yr MA & HIT $-$31d & HIT $-$42d & HIT $-$44d & no fire & no fire & no fire \\
\hline
\end{tabular}
\end{table}

\begin{figure}[htbp]
\centering
\includegraphics[width=\linewidth]{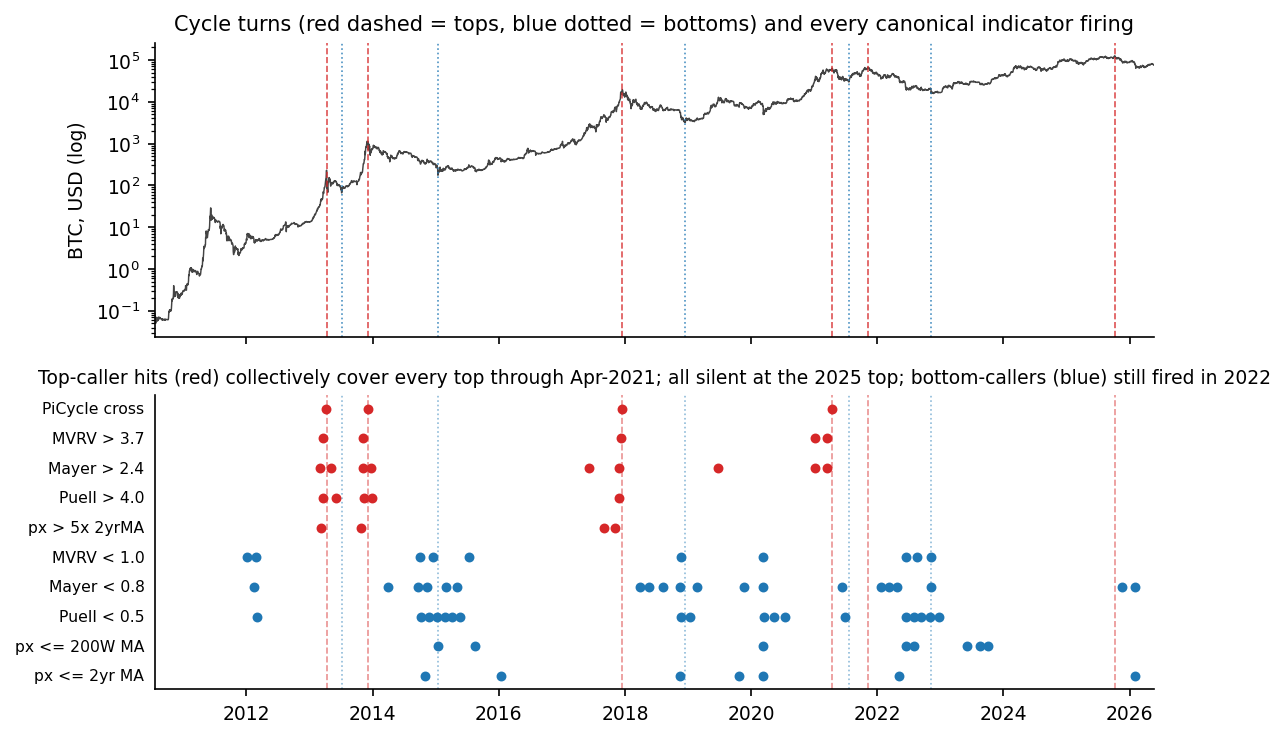}
\caption{Every canonical indicator firing, 2011 to 2026. The top-callers (red) collectively marked every top through April 2021 and were all silent at the October 2025 top; bottom-callers (blue) still fired in 2022.}
\label{fig:fig5}
\end{figure}

The degradation is ordered by threshold aggressiveness (Figure~\ref{fig:fig5}). The most-stretched thresholds (Puell 4.0, 5$\times$ the 2-year MA) fell silent after 2017; the remainder fired at the April 2021 secondary peak and missed the bear-defining November top by seven to eight months; by October 2025 every top-caller was silent. Bottom-callers lasted longer: MVRV \textless{} 1.0, Mayer \textless{} 0.8 and Puell \textless{} 0.5 all marked the November 2022 bottom within a week. Their current-cycle test is live, however, and several thresholds are unreached at the trough so far (minimum MVRV 1.15; price has not touched the 200-week MA at all this cycle). The 200-week MA itself has already failed in both directions: the December 2018 bottom formed just above it (no touch at daily closes), and 2022 broke 34\% below it.

We attribute these triggers to their originators, all practitioner constructions without peer-reviewed sources: the Pi Cycle Top to Swift (2019) \cite{swift2019}, the Mayer Multiple to Mayer (2016) \cite{mayer2016}, the Puell Multiple to Puell (2019) \cite{puell2019}, and MVRV to Mahmudov and Puell (2018) \cite{mahmudov2018}; the 200-week moving average is common trader convention. We use each threshold as its source defines it.

\subsection{The mechanism: amplitude is dying}

\textbf{Table 3. Per-epoch indicator extremes.}

\begin{table}[htbp]
\centering
\small
\begin{tabular}{lccccc}
\hline
 & Pi ratio & MVRV & Mayer & Puell & 2yr-MA mult \\
\hline
E1 max & 1.23 & 5.88 & 8.26 & 10.49 & 17.96 \\
E2 max & 1.06 & 4.72 & 3.78 & 6.62 & 9.96 \\
E3 max & 1.00 & 3.96 & 2.83 & 3.46 & 4.88 \\
E4 max & 0.74 & 2.74 & 1.52 & 1.59 & 2.31 \\
E1 min (MVRV/Puell) &  & 0.56 &  & 0.31 &  \\
E4 min (MVRV/Puell) &  & 1.15 &  & 0.53 &  \\
\hline
\end{tabular}
\end{table}

\begin{figure}[htbp]
\centering
\includegraphics[width=\linewidth]{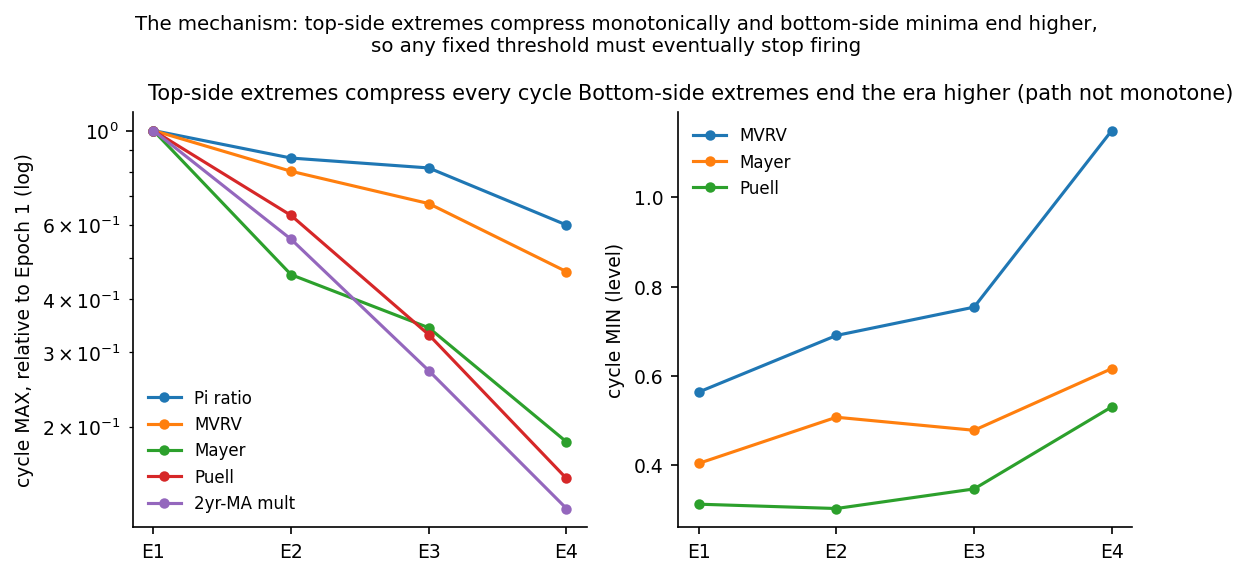}
\caption{The compression mechanism: every top-side per-cycle maximum declines monotonically; bottom-side minima end the era higher than they began, though not monotonically.}
\label{fig:fig3}
\end{figure}

Every top-side maximum in Table 3 declines monotonically across all four epochs. The bottom-side minima end the era far higher than they began (MVRV 0.56 $\rightarrow$ 1.15; Puell 0.31 $\rightarrow$ 0.53), though not monotonically (Figure~\ref{fig:fig3}). The oscillation range is narrower at the era's end from both sides. The consequence is mechanical: any fixed threshold calibrated on past cycles must eventually stop firing, in order of how deep into the old range it reaches. Table 2's death sequence is Table 3's compression, realized. "It never missed before" described a shrinking range, not a reliable signal. Recalibrating thresholds each cycle is simply a bet that the compression has stopped, fit to the past.

\subsection{Continuous indicators decay and invert; the time-trend survives}

\begin{figure}[htbp]
\centering
\includegraphics[width=\linewidth]{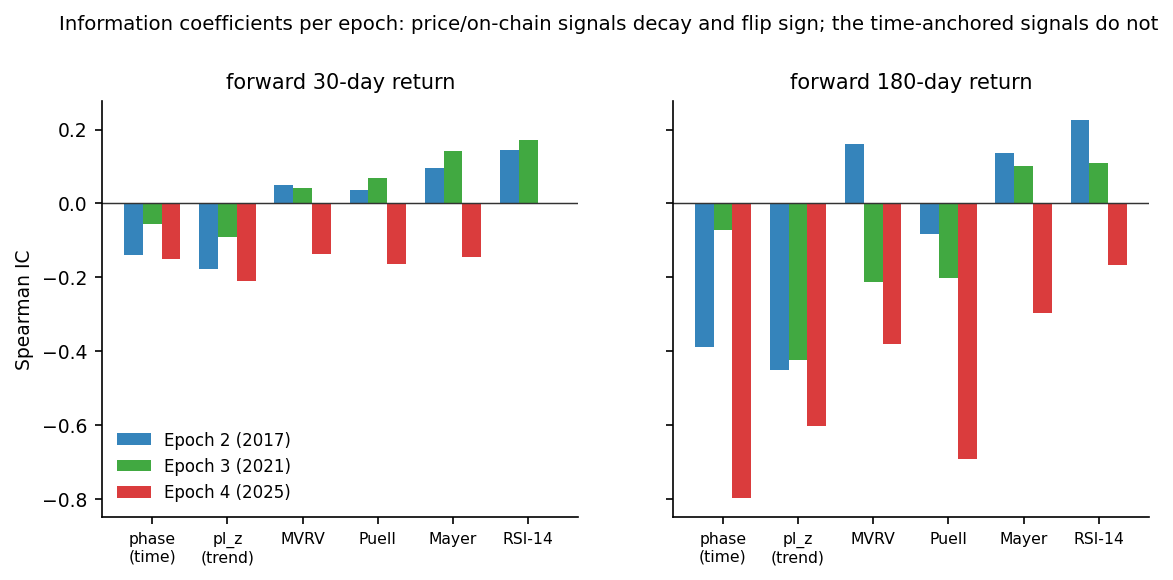}
\caption{Information coefficients per epoch (descriptive): fast price and on-chain signals decay or flip sign across cycles; the time-anchored and slowest trend-anchored signals hold direction.}
\label{fig:fig4}
\end{figure}

Thresholds aside, the underlying signals' rank correlation with forward returns tells the same story. RSI-14's 30-day IC across epochs runs +0.28 $\rightarrow$ +0.14 $\rightarrow$ +0.17 $\rightarrow$ $-$0.02: decay to statistical zero, robust to every RSI period from 7 to 30. At the 180-day horizon several indicators \emph{invert}: the Pi ratio's IC swings to +0.49 in E4, and the Mayer z and RSI flip sign relative to their 2017-era values. An indicator whose sign is era-dependent is unusable ex ante regardless of magnitude (Figure~\ref{fig:fig4}).

Against this, the power-law deviation z is the only signal in the formal inference battery (power-law z, Mayer z, RSI) whose 180-day IC has a stable sign and strong magnitude in all three mature epochs: $-$0.41, $-$0.40, $-$0.51 (Bitstamp; $-$0.45/$-$0.42/$-$0.60 on Coin Metrics). In the wider descriptive battery, the only other constructions that hold direction at this horizon are themselves slow and trend-anchored (price relative to the 200-week average; the Puell Multiple) or pure time (phase); every fast oscillator flips. Its nominal HAC statistics are large in both recent epochs (E3: t = $-$2.69; E4: t = $-$2.63), and it is the only cell in the grid with same-sign nominal significance in both. We do not advance this as calibrated inference, because the rotation null of Section 4.3 shows that configuration arises by persistence artifact with probability 0.21, and that the NW statistic's true size here is 0.17--0.33, not 0.05. What we do advance is model-free: across three mature epochs and two price sources the sign never flips and the magnitude never weakens, while every comparator flips, fades, or both; the identical pattern reproduces on a second asset with a different exponent (Section 5.5); and the economic crossover below is out of sample. The paired contrast |IC\_pl| $-$ |IC\_rsi| rises monotonically across epochs (+0.15 $\rightarrow$ +0.23 $\rightarrow$ +0.28 $\rightarrow$ +0.31), a descriptive trend subject to the same inferential caution. With at most seven or eight non-overlapping 180-day windows per epoch (three or four in the current one), this is the honest power limit of the data, and we say so rather than dress it.

\begin{figure}[htbp]
\centering
\includegraphics[width=\linewidth]{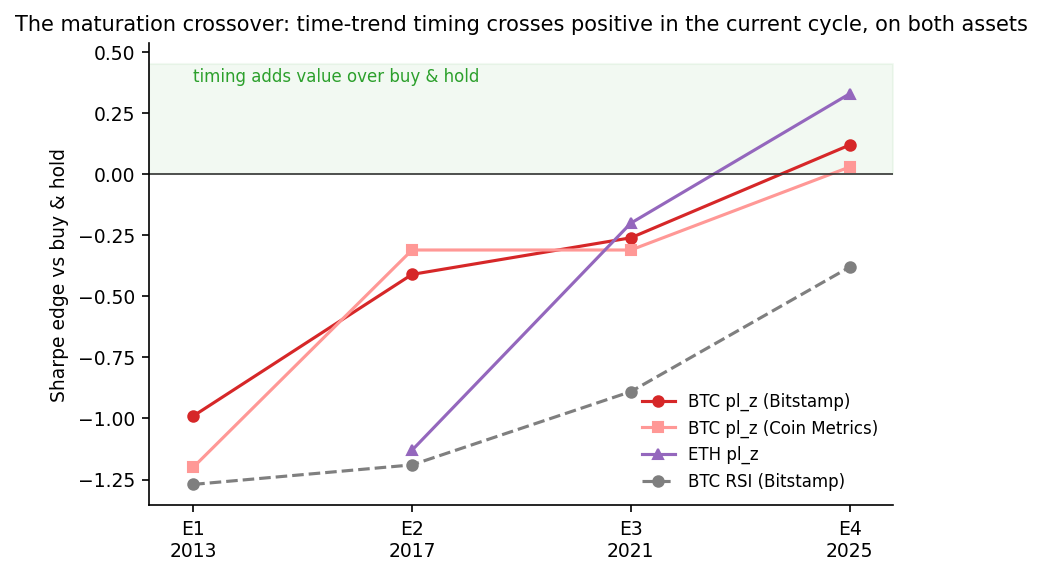}
\caption{The maturation crossover: the power-law timing rule's Sharpe edge versus buy-and-hold improves monotonically across epochs and crosses positive in the current cycle on both BTC and ETH (the 2025 to 26 holdout is out of sample; suggestive, see Section 5.4).}
\label{fig:fig6}
\end{figure}

Design (a) of Section 4.4 produces too few events to evaluate: the power-law rule fired six times in fifteen years (positive in every epoch, with two stop-outs), so the economic evidence rests on design (b). Economically, no timing signal beats buy-and-hold in the early, bull-dominated epochs: sitting flat costs upside, and we report this plainly. But the power-law signal's Sharpe edge over buy-and-hold improves monotonically across epochs ($-$0.99 $\rightarrow$ $-$0.41 $\rightarrow$ $-$0.26 $\rightarrow$ +0.12) and crosses positive in the current cycle (Figure~\ref{fig:fig6}). Out of sample (2025--26), it beats buy-and-hold on Sharpe while cutting maximum drawdown by ten points ($-$41\% versus $-$51\%), and it is the only signal that does both (the Mayer z edges buy-and-hold on Sharpe alone). The weight this can bear is limited: the holdout is a single 526-day window dominated by one bear market, so the positive crossing is suggestive rather than decisive. What carries evidential weight is the shape rather than the endpoint, a four-epoch monotone improvement reproduced on an independent price source and on a second asset with a different exponent. What would make it decisive is not more backtesting but the pre-registered windows of Section 8.

A complementary, deliberately simpler economic test points the same way while exposing the limits of all of this. A purely halving-anchored cash rule (hold Bitcoin except during the historical top-to-bottom window, roughly 525 to 900 days after each halving, using only the known halving calendar and a one-day lag) beat buy-and-hold on total compounded return in all four cycles, cut every cycle's maximum drawdown, and compounded to roughly fifty times buy-and-hold's terminal wealth over the full history. This is not generic drawdown avoidance. Sliding the same-width cash window across the cycle, the edge is sharply localized to the true bust phase and loses everything when placed on the boom; a random cash mask of identical duration beats buy-and-hold only four percent of the time and never approaches the realized multiple; and the rule survives leave-one-cycle-out, a window optimized on any three cycles still beating buy-and-hold out of sample on the fourth (ratios 1.7 to 4.1). We nonetheless decline to advance it as a forward-validated edge, and the contrast with the Sharpe result above is the reason for care, not for excitement. The two are not in tension: total return rewards sidestepping catastrophic drawdowns, while risk-adjusted Sharpe penalizes the opportunity cost of sitting in cash during a secular bull, so the same timing can win on return and lose on Sharpe in the same cycle. The fifty-fold figure is overwhelmingly a compounding artifact of avoiding four single-asset crashes; the honest per-cycle out-of-sample edge is the two-to-fourfold range; and with four cycles, leave-one-out tests consistency across the turns already observed, not robustness to the next cycle's drift. We report it as a complementary illustration that the timing carries real economic weight, not as a trading system.

The crossover requires the moderate entry threshold (z \textless{} 0 or +0.25; the deepest threshold remains slightly negative), a threshold dependence we flag rather than bury. The result is insensitive to the fit and z-score warmup parameters in the two most recent epochs (nine variants within $\pm$0.01 IC; the 2016--2020 epoch varies by up to 0.09 under the longest fit warmup) and strengthens monotonically with horizon (E4 IC $-$0.19 at 60 days, $-$0.69 at 360).

The first epoch is the exception throughout: the causal fit is still immature in 2012--13 (the causal exponent starts steep, near 12 at the first valid fit in 2011, and is still settling toward its $\approx$ 5.6 asymptote through 2013), and the deviation score's E1 sign is flipped. We treat this as a feature of the methodology rather than an embarrassment. A causal fit needs roughly one cycle of data to learn the trend, after which deviations become informative. The identical warmup artifact appears in Ethereum's first epoch (below), supporting this interpretation.

\subsection{Replication: independent source and second asset}

On the independent Coin Metrics series, the IC results reproduce slightly stronger (180-day IC reaching $-$0.60 in E4), and the Sharpe-edge sequence reproduces the same monotone improvement with a weaker positive endpoint ($-$1.20 $\rightarrow$ $-$0.31 $\rightarrow$ $-$0.31 $\rightarrow$ +0.03). On Ethereum, with its own genesis (July 30, 2015) and a causally fit exponent near 2.1, very different from Bitcoin's 5.6, the same maturation crossover appears: 180-day IC strengthening from $-$0.37 (E3) to $-$0.65 (E4), and a Sharpe edge over buy-and-hold of $-$1.13 $\rightarrow$ $-$0.20 $\rightarrow$ +0.33 across the market-cycle epochs, while RSI never goes positive in any epoch. Ethereum's first epoch shows the same warmup sign-flip as Bitcoin's. That the crossover replicates on a second asset with a different exponent indicates a property of maturing crypto assets, not a Bitcoin-specific curve. The alignment runs deeper than the trend: Ethereum's mechanical cycle \emph{tops} themselves land near the BTC halving phase, at 553, 546, and 489 days after the BTC halving (against Bitcoin's 525/546/534), with the 2021 tops falling on the same day. A coin with no halving of its own still turns near the BTC halving phase, which points to the halving as a market-wide clock rather than only a Bitcoin-supply effect.

\subsection{The macro alternative: era-dependent and ex post}

Two confounders dominate popular discussion: global liquidity (money supply) and the business cycle. We state the identification problem plainly: the liquidity cycle and the halving cycle are roughly co-periodic, and with three to four overlaps no co-periodic confounder can be fully excluded. Four discriminating tests are nonetheless available, and all four point the same way.

\textbf{(i) Sign stability.} M2 YoY growth's 180-day IC with forward Bitcoin returns runs +0.41 (E2), $-$0.10 (E3), $-$0.83 (E4). The yield-curve spread runs +0.42, $-$0.35, $-$0.79. Both flip sign across epochs exactly as the price oscillators do. The E4 sign is backwards for the liquidity narrative: from the April 2024 halving to the data freeze, M2 expanded continuously (YoY growth between +0.9\% and +4.7\%, with no contraction) while Bitcoin topped and fell.

\textbf{(ii) Lead/lag stability.} The folklore that "M2 leads Bitcoin by about 10 weeks" is a one-era artifact. The IC of lagged M2 growth against forward 90-day returns is positive at every lead from 0 to 270 days in E2 (peaking near +0.45 at a 120-day lead, presumably the folklore's origin), negative throughout E3, and negative at every lead in E4. No stable lead exists.

\begin{figure}[htbp]
\centering
\includegraphics[width=\linewidth]{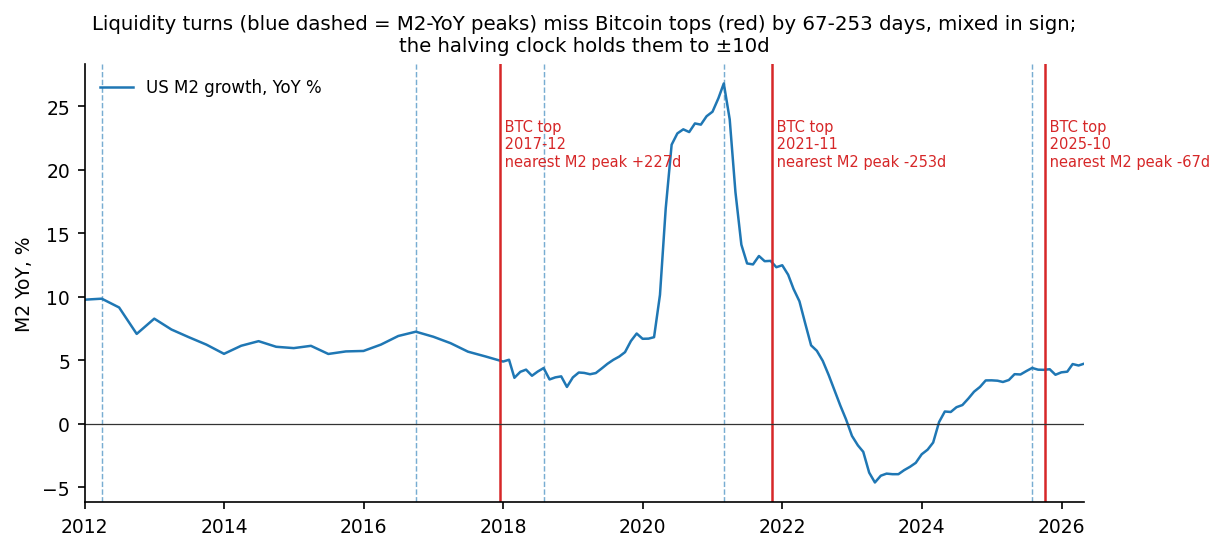}
\caption{Liquidity turns versus the clock: M2-growth turning points miss Bitcoin tops by 67 to 253 days with mixed sign; the halving clock holds them to plus or minus 10 days.}
\label{fig:fig7}
\end{figure}

\textbf{(iii) Turn-timing precision.} Distances from the three Bitcoin tops to the nearest M2-YoY peak are +227, $-$253, and $-$67 days: mixed in sign and ranging from 67 to 253 days, versus the halving clock's $\pm$10 days (Figure~\ref{fig:fig7}), an order of magnitude looser. There is also a practical asymmetry: monetary turning points are confirmable only months after the fact, while the halving schedule has been public since 2009. We credit the macro side its successes: the February 2021 M2-growth peak preceded the April 2021 secondary top by six weeks, and the 2022 curve inversion coincided with that bear market.

\textbf{(iv) Horse race.} Regressing forward 180-day returns on standardized phase, M2 growth, and the yield spread jointly (Newey--West, lag 180, mature sample, n = 3,426) gives phase t = $-$2.00, M2 t = +0.20, spread t = $-$0.07. Controlling for one another, only the clock retains any weight. The persistent-regressor calibration caveat of Section 4.3 applies to these t-statistics with full force, so we read the horse race as a relative statement (which regressor survives joint conditioning with stable direction) rather than as calibrated significance. Per-epoch partials are noisy under collinearity (|$\rho$| $\approx$ 0.5--0.66) and we do not over-read them.

\begin{figure}[htbp]
\centering
\includegraphics[width=\linewidth]{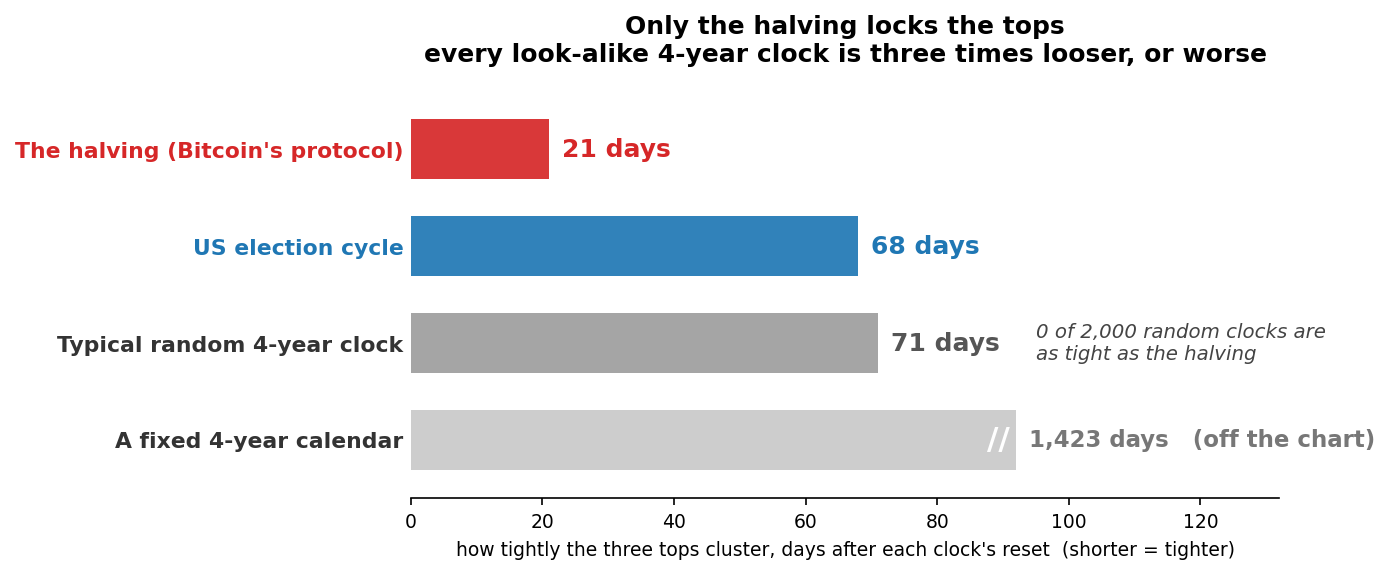}
\caption{The four-year-clock placebo. The three mature tops cluster within 21 days measured from the halving, versus 68 days from the US election cycle, 71 days from a typical random four-year clock, and 1,423 days from a fixed four-year calendar. None of 2,000 random four-year clocks is as tight as the halving, so the regularity is specific to the halving event, not a generic four-year rhythm.}
\label{fig:figB}
\end{figure}

\textbf{(v) A four-year-clock placebo.} The halving is one roughly-four-year clock; is the regularity halving-specific, or would any four-year cycle do as well? Measured under the halving, the three mature tops span 21 days. Under the US presidential-election cycle, also four-yearly, they span 68 days; under a fixed four-year calendar they span 1,423 days, because the tops follow the halving's actual, lengthening spacing rather than a fixed period; and of 2,000 random four-year clocks, \emph{none} matches the halving's 21-day spread. The clustering is specific to the halving event, not to the calendar, which is the cleanest separation we can draw between the protocol clock and any generic four-year macro or political rhythm (Figure~\ref{fig:figB}).

The conclusion is not that macro is irrelevant; within any single era its correlations are large. It is that macro's relationship to Bitcoin is era-dependent and ex post, sharing the instability of every price-anchored signal, while the time structure is era-stable and ex ante. In the one episode where the two clocks disagreed (late 2025, liquidity expanding, the calendar striking day 534), price followed the code.

\subsection{The Satoshi Clock}

The framework condenses to a two-coordinate state description we call the \textbf{Satoshi Clock}:

\begin{itemize}
\item \textbf{CLOCK}: days since the most recent halving, the angle of the four-year revolution;
\item \textbf{SPRING}: the causal power-law deviation z, the radius, how stretched price is from its time trend.
\end{itemize}

\textbf{Table 4. The damped spiral (SPRING at each turn).}

\begin{table}[htbp]
\centering
\small
\begin{tabular}{lcc}
\hline
Turn & CLOCK (days) & SPRING (z) \\
\hline
Top 2013 & 371 & +2.85 \\
Bottom 2015 & 777 & $-$1.46 \\
Top 2017 & 525 & +2.69 \\
Bottom 2018 & 889 & $-$0.61 \\
Top 2021 & 546 & +1.29 \\
Bottom 2022 & 912 & $-$1.17 \\
Top 2025 & 534 & +0.43 \\
\hline
\end{tabular}
\end{table}

\begin{figure}[htbp]
\centering
\includegraphics[width=\linewidth]{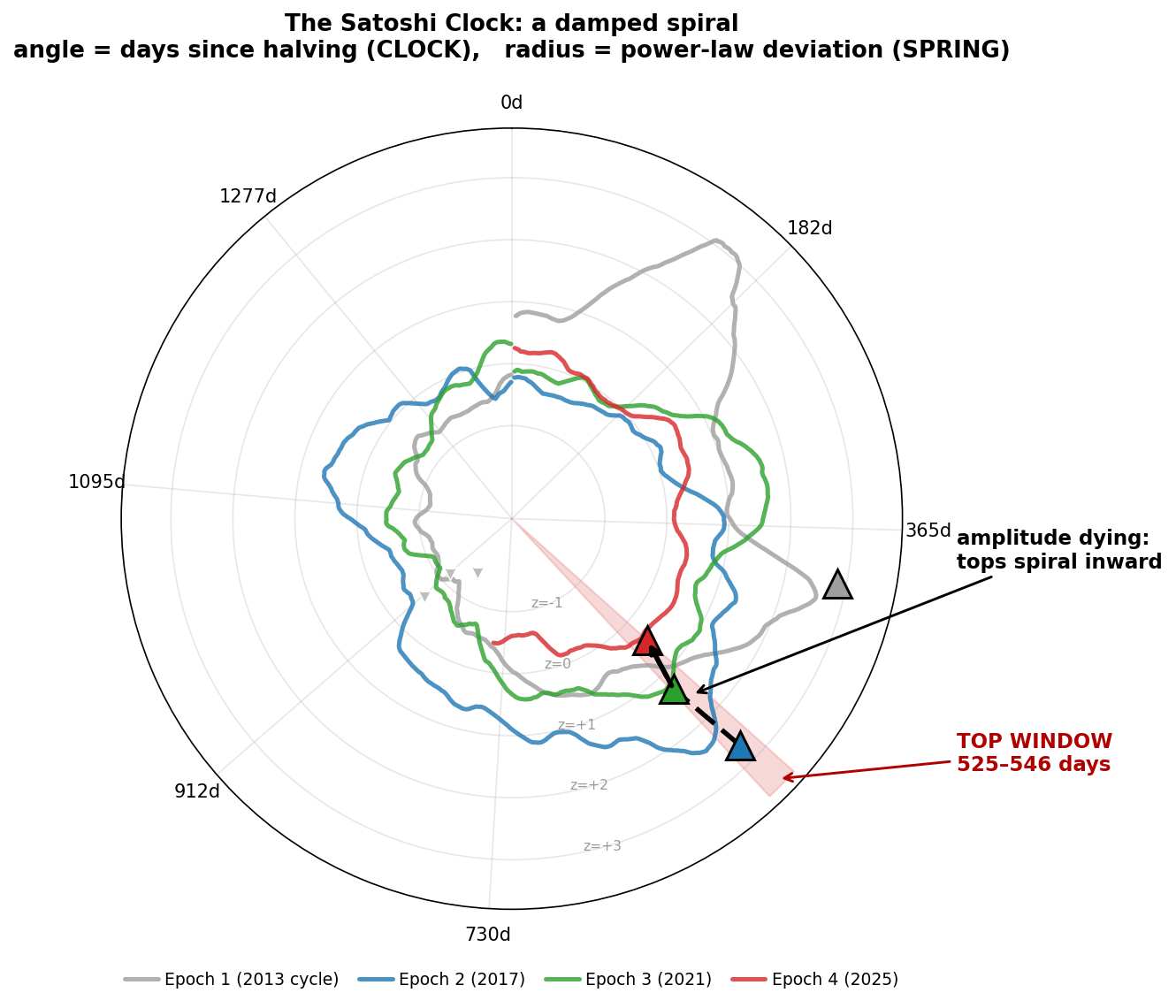}
\caption{The Satoshi Clock: Bitcoin's full history as a damped spiral. Angle = days since halving (CLOCK); radius = causal power-law deviation (SPRING). The three mature tops (triangles) all fall in the shaded 525 to 546-day wedge, while their radius shrinks each cycle along the dashed inward arrow (SPRING +2.69 then +1.29 then +0.43). The grey loop is the immature first cycle (2013, top at 371 days). The amplitude is dying; the clock is not.}
\label{fig:fig2}
\end{figure}

Read down Table 4's SPRING column at the tops: +2.85 $\rightarrow$ +2.69 $\rightarrow$ +1.29 $\rightarrow$ +0.43. At the tops, the radius shrinks every revolution (the bottom-side readings are noisier: $-$1.46, $-$0.61, $-$1.17). This is Table 3's compression in trend-deviation units, and it is why every amplitude-calibrated indicator has died or will die. Read down the CLOCK column at the mature tops: 525, 546, 534. The angle does not move. \emph{The amplitude is dying; the clock is not.} Bitcoin's full price history, plotted in these coordinates, is a damped spiral converging toward its power-law trend (Figure~\ref{fig:fig2}). The spiral is also a quantitative statement of the asset's maturation: each cycle, Bitcoin overshoots its own adoption trend by less.

As of the Coin Metrics data freeze (May 23, 2026; price \$76,620): CLOCK 763, SPRING $-$0.53, 229 days after the October 2025 top. Mid-bear, by the geometry of every prior cycle.

\subsection{The clock governs the whole cycle, not just the turns}

\begin{figure}[htbp]
\centering
\includegraphics[width=\linewidth]{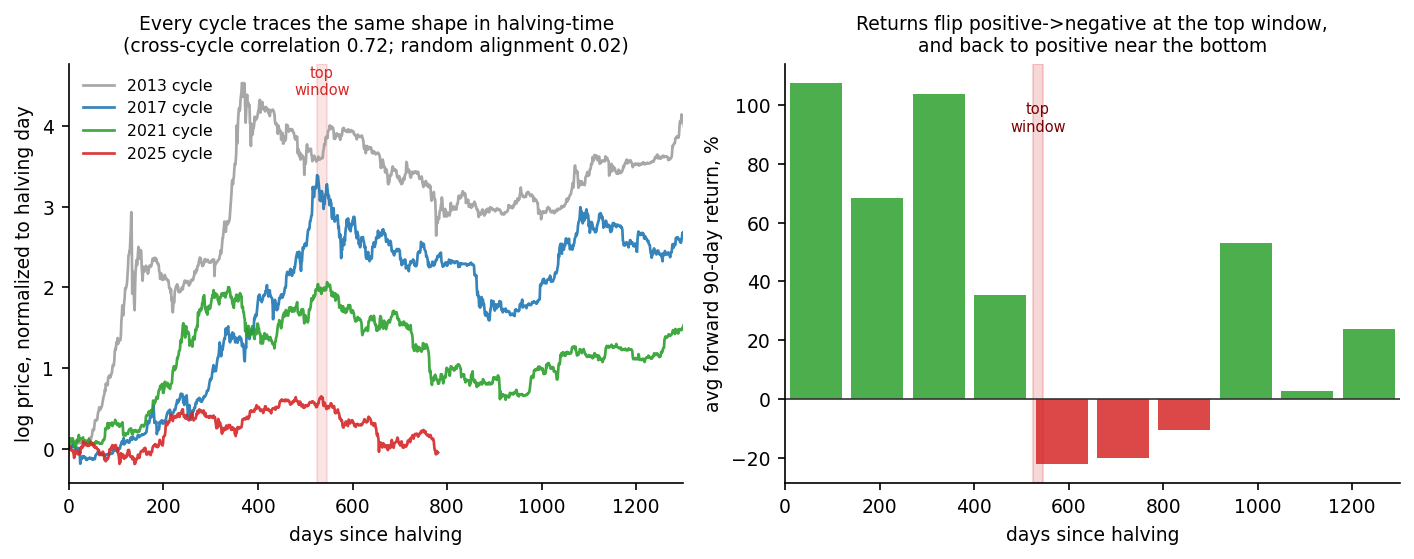}
\caption{The clock governs the whole cycle, not just the turns. Left: each cycle's price path aligned by days since halving (cross-cycle correlation 0.72; random alignment 0.02). Right: average forward 90-day return by phase, flipping positive to negative at the mechanically dated top window (525 to 546 days) and back near the bottom.}
\label{fig:fig8}
\end{figure}

The turn-dating tests treat each cycle as two points. Two further tests show the clock organizes the entire trajectory. First, when the three completed cycles are aligned by days since halving and normalized to the halving-day price, their paths correlate at 0.72 on average: the boom, the top, the bust, and the recovery all line up (Figure~\ref{fig:fig8}, left). Aligned at a random offset instead, the correlation collapses to 0.02, and only 2.8\% of random alignments match the halving alignment, so it is the halving phase specifically that the cycles share, not merely a smooth-curve artifact. Second, pooling every (phase, forward-90-day-return) pair across cycles, average forward returns are strongly positive from the halving to about day 520, turn negative from roughly day 520 to 900, and turn positive again afterward (Figure~\ref{fig:fig8}, right). The sign flips almost exactly at the mechanically dated top window (525--546 days) and again near the bottom. Turn-dating and average returns, two independent methods, locate the same turn days.

Not everything locks to the clock, and we report the miss because it disciplines the hits. Realized volatility does \emph{not} peak at a consistent halving-day: the peak phases are 157, 585, and 394 days, and the volatility-versus-phase shape correlates at only 0.15 across cycles, indistinguishable from the random-offset placebo (p = 0.19). The clock governs the cycle's direction and shape, not its volatility. A framework that found a clock in everything would be the more suspicious one; this one finds the clock where there is real structure and reports plainly where there is not. The placebo probabilities in this section (the 2.8\% above and the volatility p = 0.19) come from random-offset permutation nulls of the same discrete-event family exempted in Section 4.3, not the autocorrelation-robust per-signal p-values that our rotation null disqualifies; they ask only how often a random phase alignment matches the halving's, which block reshuffling answers without the persistence bias that voids HAC inference here.

\subsection{The clock in block height, its native unit}

The halving is defined in blocks, one every 210,000, not in days, and the chain does not run at a fixed 144 blocks per day, so "days since halving" and "blocks since halving" are two different clocks. Because the date-to-height map (Section 3) is fixed by the chain and independent of price, the turn-timing battery re-scores in the protocol's own coordinate without circularity, on the height-covered span ending May 23, 2026. We report gains and losses in full.

\emph{The tops hold, and their separation from look-alike clocks sharpens.} In block phase the deterministic top construction is reproduced by zero of 10,000 block-bootstrap paths, exactly as in calendar days; the conservative selection-symmetric construction falls from 11 to 3 per 10,000. The three mature tops sit 77,901 to 79,596 blocks after their halvings (a 1,695-block band by the mempool point-lookups, 1,723 by the mechanical dating on the daily height series, agreeing within end-of-day sampling). The placebo behaves in a revealing way. In days, a fixed four-year clock at a random phase never matches the halving's top cluster, so phase carries the signal; in blocks, a fixed 210,000-block clock at a random offset matches it 98.5\% of the time, because the halving is an exact period, not a drifting one, and the tops recur on it regardless of phase. Draw the period at random and only 0.9\% match. The halving out-separates its look-alikes more widely than in days, a 1,723-block band against 12,775 for the US election clock and 209,158 for a fixed calendar. The bottom-anchor question of Section 4.5 is thereby resolved mechanically: measuring a block-defined cycle in days created it, and in blocks it dissolves.

\emph{The bottoms are partially, not fully, rescued.} The paper's demoted bottom statistic stays demoted. The top-to-bottom gap that Section 4.5 found process-intrinsic is reproduced by about 40\% of null paths, unchanged, so bear duration still carries no clock evidence. What the block coordinate adds is a phase statistic the calendar could not resolve: the three bottoms' blocks-after-halving span is reproduced by 1.7\% of paths in day phase and only 0.16\% in block phase, a tenfold sharpening that survives conditioning on the cycle structure (3.7\% to 0.3\%). An adversarial check locates the effect precisely. The top-to-bottom lag across the three cycles is not itself sharpened by block coordinates (0.42\% in days, 1.11\% in blocks), so the bottoms are not merely inheriting the tops' phase-lock; each turn is independently anchored to the halving, tops and bottoms occupying separate tight bands near 78,000 and 132,000 blocks. We read this as a partial rescue on three cycles, a weak prior worth watching, not a reversal of the demotion.

\emph{The 2013 top partially reconciles.} The first-cycle top sits at 0.693 of the mature mean in days but 0.801 in blocks, because the 2011 to 2013 chain ran about 15\% fast (170 against 147 blocks per day at maturity); roughly a third of its apparent earliness is the young chain outrunning the calendar. Consistent with this, the hostile four-top variant that forces the 2013 top into the mature set widens the null from zero to 8 per 10,000 in days but only to 3 per 10,000 in blocks: the native coordinate is more robust to the first cycle.

\emph{What does not improve, stated as plainly as what does.} The cycle-shape overlay of Section 5.8 is marginally worse in block phase (mean cross-cycle correlation 0.69 against 0.72 in days), and the volatility clock, already the section's honest miss, is worse still (placebo p = 0.40 against 0.19). The block clock times the cycle's discrete turns, including the return-by-phase sign flip, which lands on the 77,901 to 79,596-block top band; it does not organize the continuous path between the turns any better than the calendar does. On balance the native coordinate strengthens the tops result's robustness and its separation from look-alike clocks, dissolves the bottom-anchor puzzle, and recovers partial bottom structure, at the cost of the shape and volatility overlays. The case still rests on the tops, and the phase-lock, not the cadence, is still the surprise.

\section{Discussion}

\textbf{Why price indicators must fail.} Under the damped-oscillation reading, the failure of the cycle-caller family is not a coincidence of 2025 but a structural inevitability with a predictable order. A threshold indicator is a horizontal line drawn through an oscillation; if the oscillation's envelope contracts monotonically, the line is exited permanently, deepest lines first. That is precisely the observed death sequence (Puell 4.0 and 5$\times$2yr-MA after 2017; Pi Cycle, MVRV 3.7 and Mayer 2.4 after April 2021; everything by 2025). Continuous versions of the same signals escape the threshold problem but not the deeper one: their \emph{meaning} is era-dependent, decaying and inverting as the market matures. The Adaptive Markets view organizes this naturally. What capital arbitrages away is the exploitable amplitude; the calendar, being no one's inefficiency, is not arbitraged but if anything reinforced.

\textbf{Why the clock might work.} We deliberately bracket the causal channel, because the two leading candidates are observationally similar at n = 3 and both leave our claim intact. (a) \emph{Supply economics:} the halving instantaneously halves the flow of new supply; with demand continuity, the price response propagates with lags set by miner inventory behavior and reflexive momentum, a mechanical clock. (b) \emph{Coordination:} the schedule is the most public focal point in the asset's culture; if enough participants expect cycle behavior in halving-time, front-running and herding can synchronize the turn, a Schelling clock (Schelling, 1960 \cite{schelling1960}). Explanation (b) is not deflationary: a self-fulfilling cycle anchored to a constant in the source code is still a cycle anchored to the source code, and its timing is still knowable decades ahead. Distinguishing (a) from (b), for instance via miner flow data or cross-asset placebo clocks, is the natural next study.

\textbf{The maturation synthesis.} The two halves of the paper are one phenomenon. Diminishing amplitude (returns, volatility, indicator extremes, SPRING at tops) is the asset converging toward its adoption trend; fixed period is the trend's clock. On this view the often-asked question "is the four-year cycle dead?" conflates the two. The cycle's \emph{profitability} is dying on schedule with maturation, while its \emph{timing} has never been more precise. We would expect future cycles to continue this pattern, shallower oscillations and punctual turns, until amplitude is no longer distinguishable from noise.

\section{Limitations}

\begin{enumerate}
\item \textbf{Retrospective identification and post-hoc testing.} The six turns were identified ex post by a mechanical rule, not called in real time; the null test was designed after the pattern was observed; and the headline specification excludes the first cycle's top while including its bottom. We disclose all three (Section 4.5), report the hostile variant (joint p up to $10^{-3}$), and emphasize that the retrospective probabilities motivate the prospective tests of Section 8 but cannot substitute for them. One verified real-time antecedent exists (Section 8.1: a public, platform-timestamped month-precision call of the 2025 top, made January 2025). It is n = 1, it is the author's own, and it applied the community's folk heuristic rather than this paper's model: an out-of-sample application of a known rule, not independent validation. We weight it accordingly.
\item \textbf{Few cycles.} The headline regularity rests on three mature tops and three completed bottoms. The variant grid (Section 5.1) shows this regularity is very unlikely under a timing-free null, but n = 3 pairs cannot establish causation, fine structure, or stationarity of the period. Every competing cycle theory faces the same constraint on the same data. We also probe the clock from several angles (top phase, bottom phase, the cycle's shape, its return profile, its volatility, and a battery of look-alike clocks); most survive their own placebo and one, volatility, does not, but at three to four cycles the battery as a whole carries a multiple-comparisons burden we cannot fully discharge. The placebos and nulls bound how surprised we should be on each test; the pre-registered predictions of Section 8 are what ultimately settle it.
\item \textbf{Per-epoch significance is unattainable here, for anyone.} No individual cell of the 36-cell grid survives false-discovery control, and our rotation null shows the HAC machinery's true size on persistent signals at these sample sizes is 0.17--0.33 against a nominal 0.05. The apparent two-epoch significance of the power-law cell arises by persistence artifact with probability 0.21. The power-law claims in this paper are therefore descriptive (sign stability, monotone trends) and economic (out-of-sample crossover), supported by cross-source and cross-asset replication, never inferential at the per-epoch level. The same caution applies to the macro horse race of Section 5.6, whose regressors are equally persistent; we read it as a direction-stability comparison, not calibrated inference.
\item \textbf{Co-periodic confounders.} Liquidity and halving cycles overlap too few times to be fully separable; our four tests discriminate on stability, precision, and the 2024--26 disagreement episode, not on exclusion.
\item \textbf{The first epoch.} The causal fit needs about one cycle of warmup; E1 results are sign-flipped and excluded from mature-epoch claims by stated rule, not post hoc.
\item \textbf{Intra-cycle structure.} The 2020--24 epoch contained a double top (April/November 2021) and an unusually early bottom relative to halving-time; phase's within-cycle IC was weak in that epoch ($-$0.07). The clock claim concerns the bear-defining turns, not every intermediate swing.
\item \textbf{Threshold dependence and a thin holdout.} The exposure-level Sharpe crossover requires the moderate entry threshold (the deepest variant remains slightly negative in E4), and the out-of-sample window is a single 526-day span dominated by one bear market: suggestive, not decisive. The claim's weight rests on the four-epoch monotone trend and its cross-source and cross-asset reproduction, pending the prospective test.
\item \textbf{Marginal head-to-head contrast.} The paired |IC| advantage of the power-law signal over RSI in E4 is significant only at the 10\% level (p = 0.07).
\item \textbf{Macro data.} US rather than global M2; quarterly macro granularity before 2018; and ex-post dating of monetary turns, which biases toward the macro alternative.
\item \textbf{Bottom-callers not yet failed.} The on-chain bottom indicators nailed 2015, 2018, and 2022, and their E4 test is in progress. If the current bear bottoms below MVRV 1.0 and at the 200-week MA, the bottom-side compression claim weakens (the top-side claim is unaffected).
\item \textbf{The block-coordinate re-analysis (Section 5.9).} The block clock shares the three-cycle limit of everything else here; its bottom rescue in particular rests on three bottoms and is a weak prior, not a result. Two guards are worth stating. First, because it compares two range statistics measured on the same three turns, a spuriously tighter block spread could in principle be two drifting series canceling; we mitigate this by scoring the null in the identical coordinate, so uniform compression cancels and only differential tightness registers, and by resting the block case on mechanism (the exact 210,000-block period) rather than on any single tighter number. Second, the height series lags the price data by about eight weeks, so the block battery runs on the span ending May 23, 2026; no turn falls in the gap, and the day-phase null re-run on the same truncated span still reproduces the zero-of-10,000 tops result, confirming the truncation is immaterial.
\end{enumerate}

\section{Pre-Registered Predictions and a Real-Time Antecedent}

\subsection{A real-time antecedent}

Two public, platform-timestamped artifacts show the timing heuristic in use nine months before the cycle top, predating all of this paper's analysis. On January 6, 2025, with Bitcoin at \$102,180, the author published a 57-second video to Instagram (account \emph{bitcoin.daily}; upload timestamp verified from platform metadata), stating: \emph{"Every crypto bull run has a peak. We saw it in 2013, right before an 86\% crash. We saw it in 2017, right before an 81\% crash, and once again in 2021, before a 77\% crash... Bitcoin follows a predictable four-year cycle based on its halving. Based on the latest halving back in April of 2024, data tells us that Bitcoin is most likely a peak in late 2025... set an alert on your phone for October of 2025."} The video includes an on-screen calendar graphic showing an all-day event for Monday, October 20, 2025, titled "SELL BITCOIN (18 Months POST HALVING)." Five days later (January 11, 2025, Bitcoin at \$94,607), a longer YouTube video on the channel "Bitcoin Daily" restated the call: \emph{"every single bull market, Bitcoin peaks 18 months after its halving... that puts the next peak in October of this year,"} and \emph{"the cycle isn't up for debate. It's literally written into its code."} Quotations are from automated transcription of the audio; punctuation is editorial, obvious captioning errors are corrected (the auto-captioner renders "halving" as "Halen"), and the spoken wording is otherwise preserved.

\textbf{Provenance, stated precisely.} The January 2025 dating of these artifacts rests on the platforms' own upload metadata, which is credible but platform-controlled; no independent timestamp of the upload dates exists. It is corroborated by engagement that accrued on the posts before the event: the YouTube video carries viewer comments that the platform dates to roughly mid-2025, months before the top, including one asking why the strategy calls for selling "around October/November," that is, third-party discussion of the timing call while the outcome was still unknown. These comment dates are platform-rendered relative dates and are approximate. No pre-event web-archive capture of either URL exists; we checked the Wayback Machine's index directly. The preservation measures taken in June 2026 (Internet Archive and archive.today snapshots, plus an OpenTimestamps anchor of the evidence bundle's SHA-256 manifest in the Bitcoin blockchain) establish only that the artifacts existed in their present form as of June 11, 2026. They protect the record against future alteration, carry no weight whatsoever on the January dating, and we claim none from them.

The realized cycle top, dated mechanically in this paper from data ending June 2026, was October 6, 2025: month-exact to both calls, fourteen days before the on-screen alert date. One further detail belongs here, since a careful reader would find it anyway. The January videos themselves recite the diminishing-crash sequence (86\%, 81\%, 77\%), the very amplitude-compression series this paper formalizes, and then extrapolate it forward as an "80 to 90\%" crash prediction, repeating at the amplitude level precisely the calibrated-on-the-past error this paper documents in every threshold indicator. The timing half of those videos has been vindicated; the amplitude half is, so far, missing high. The forecaster enacted the paper's thesis in both directions at once.

Full-record disclosure, because a forecast only counts in the context of all forecasts made: (i) the same video predicted a post-peak drawdown of "80 to 90\%," an amplitude claim that is currently overshooting (maximum drawdown near 50\% to date), consistent with this paper's compression finding and a miss in its own right if the bear ends near current levels; (ii) in other public content during the same cycle the author issued price targets of \$150,000 to \$200,000 for the top, which missed (the cycle's intraday all-time high was \$126,296); (iii) the author also published a series of 2026 bear-market leg forecasts based on cross-cycle pattern correlation, which do not derive from the clock and are claimed as no evidence here. The author operates a media channel advocating this framework, a conflict we disclose.

We weight this antecedent honestly, beginning with what it is not. The 18-months-after-the-halving rule is common knowledge in the practitioner community, not this paper's invention. What this paper contributes is the rigorous version: the measured 525/546/534-day cluster, the nulls it survives, and the compression mechanism that explains why timing outlives every other signal. The January videos are therefore an out-of-sample application of a known heuristic, publicly committed to before the fact and correct on timing. They are evidence that the regularity was real and exploitable ex ante, not independent proof of the model developed here, which did not yet exist. Within those bounds it is n = 1, it is the author's own call, and naming one month roughly nine months ahead carries non-trivial luck probability. We include it for two reasons. First, it establishes that the timing regularity was being acted on publicly, in real time, before any of this paper's analysis existed; the retrospective results of Section 5.1 did not manufacture it. Second, the bifurcation within the author's own record (the timing call hit to the month while every amplitude call missed high) is the paper's thesis enacted in miniature, by a forecaster who did not yet have the formal result. The clock is predictable; the amplitude is not.

\subsection{Pre-registered windows}

A theory of timing should stake dates, especially one whose historical evidence is otherwise retrospective. The windows below are stated before the outcome exists, and they are the standard by which the thesis should ultimately be judged. From the clock's geometry, fixed before the fact:

\begin{enumerate}
\item \textbf{Current-cycle bottom window: October 5 to November 16, 2026} (364--406 days after the October 6, 2025 top). A second anchor sharpens it: measured backward from the next halving, the interval at which past bottoms cluster most tightly (Section 4.5), the window is October 21 to November 19, 2026 (provisional, since the next halving's date drifts), so the two anchors overlap on roughly October 21 to November 16, the robust core of the call. We also register a \emph{structural} sub-prediction that falls out of Section 4.5: the 2026 bottom should sit tighter to the next-halving anchor than to the after-top anchor. That is the forward test of an anchor we could only select in-sample, and it is the honest way to earn back what the after-top measurement gave away. We still treat the bottom as the weaker of our two predictions, since bear-duration clustering from the top is substantially process-intrinsic; a hit is modest evidence, a miss modest counter-evidence. In block coordinates the same bottom maps to a height band of 968,910 to 973,934 (128,910 to 133,934 blocks after the April 2024 halving at height 840,000), a supplementary and sharper-unit statement of the same call, computed and logged on July 21, 2026, before the window opens. The height target is fixed by protocol arithmetic; only its calendar arrival depends on realized block times.
\item \textbf{Next cycle top window: 525 to 546 days after the next halving.} We state this relative to the halving rather than as a fixed calendar date on purpose. The halving's exact day is not yet set: it depends on realized block times and has historically arrived somewhat before the naive ten-minute estimate, so a calendar window pinned now would inherit an error we do not control and cannot honestly pre-register. At the current halving estimate of roughly April 2028 the window maps to approximately late September to mid-October 2029, a provisional range we will refine once the halving block is in sight. The falsifiable claim is the halving-relative one. This is the sharp test: top-phase alignment is what the deterministic null never reproduced in 10,000 paths (and the conservative variant reproduced in fewer than 0.25\% of them), and it is where the thesis lives or dies. In block coordinates this window is a height band of 1,127,901 to 1,129,596 (77,901 to 79,596 blocks after the next halving at height 1,050,000), also computed and logged July 21, 2026. This is the block clock's specific advantage here: the height target needs no estimate of the halving's uncertain calendar date, which is the very source of imprecision the calendar version cannot avoid.
\item \textbf{Falsification:} either pre-registered turn missing its window by more than about 90 days, that is the 2026 bottom falling outside roughly July to December 2026 (a daily-close cycle low more than 90 days from the October 21 to November 16 core) or the next top landing more than 90 days off 525--546 days after its halving, or a SPRING amplitude at the next top \emph{exceeding} the 2025 value (+0.43), would each contradict the damped-clock model as stated.
\end{enumerate}

We commit to evaluating these in a follow-up regardless of outcome.

\section{Conclusion}

Every price-anchored description of Bitcoin's market cycle is dying, in a measured, monotone, mechanistically explicable way: the oscillation's amplitude compresses every cycle, so thresholds go silent in order of aggressiveness and continuous indicators decay toward noise or invert. The macro alternatives behave no better, flipping sign era to era and failing a joint test against the calendar. What remains is the one structure that was never estimated from price at all, because it was never estimated from anything: a constant written into the protocol at block zero and announced by its author in January 2009 as "the amount cut in half every 4 years." In the coordinates of that clock, the asset's history is a damped spiral, with turns at 525--546 days after each halving and 364--406 days after each top, a retrospective regularity whose chance probability ranges from $5\times 10^{-6}$ to $1\times 10^{-3}$ across every timing-free null and construction we tested against it, including a 2025 top that arrived on schedule while every famous indicator was silent and global liquidity argued the other way. The amplitude is dying; the clock is not. Bitcoin, at the end of the day, is code. On the evidence assembled here, it keeps its own time.

\section*{Appendix A: Data and Code Availability}

All data are from public sources (Bitstamp public API; Coin Metrics community data; the FRED series M2SL and T10Y2Y, obtained via the TradingView mirror of the FRED catalogue with values embedded and internally cross-checked in \texttt{build\_macro\_csv.py}). The full analysis pipeline is available as documented Python scripts: \texttt{pull\_btc\_daily.py}, \texttt{pull\_btc\_ohlc.py}, \texttt{pull\_btc\_metrics.py}, \texttt{pull\_eth\_daily.py}, \texttt{build\_macro\_csv.py} (data); \texttt{pl\_lib.py} (causal power law and signal library); \texttt{study\_signals.py} (IC framework); \texttt{ic\_significance.py} (HAC and bootstrap inference); \texttt{pl\_backtest.py}, \texttt{exposure\_study.py} (economic evaluation); \texttt{replicate.py} (cross-source and Ethereum replication); \texttt{sweep\_sensitivity.py} (robustness grids); \texttt{cycle\_callers.py} (threshold indicators, extremes, turn dating); \texttt{macro\_study.py} (confounders); \texttt{satoshi\_clock.py} (permutation test and the indicator); \texttt{verify\_hardening.py} (permutation variant grid and BH-FDR across the full inference grid); \texttt{joint\_replication\_test.py} (the dependence-preserving rotation null and NW calibration diagnostic); \texttt{empirical\_null\_test.py} (the block-bootstrap drawdown-process null for turn timing, both constructions); \texttt{clock\_explorations.py}, \texttt{clock\_explorations2.py} (the four-year-clock placebo, cross-asset turn alignment, cycle-shape overlay, return-by-phase, and volatility-clock null of Sections 5.5--5.8); \texttt{clock\_anchors\_test.py}, \texttt{clock\_bottom\_anchor\_null.py} (the three-anchor bottom analysis of Section 4.5); \texttt{build\_height\_series.py} (the daily date-to-height series and its ground-truth validation); \texttt{block\_phase\_null.py}, \texttt{block\_phase\_tests.py} (the block-coordinate re-run of the turn-timing null, the placebo clocks, and the cycle-shape, return-by-phase, and volatility overlays of Section 5.9). Every numerical claim in this paper maps to a line of script output; the \texttt{audit/} directory contains the captured output of an end-to-end re-run of the entire pipeline, and \texttt{RESULTS\_LOCK.md} indexes each claim to its script and output file. A public repository link will accompany submission.

\nocite{*}
\bibliographystyle{plainnat}
\bibliography{references}
\end{document}